\documentclass[aps,twocolumn,superscriptaddress]{revtex4}

\usepackage{graphicx}
\usepackage{dcolumn}
\usepackage{bm}
\usepackage{amsmath,amssymb}
\usepackage[dvips]{color}

\def\be{\begin{equation}}
\def\ee{\end{equation}}
\def\e#1{\label{#1}\end{equation}}
\def\bea{\begin{eqnarray}}
\def\eea{\end{eqnarray}}
\def\ea#1{\label{#1}\end{eqnarray}}

\def\bem#1{\begin{mathletters}\label{#1}}
\def\eml{\end{mathletters}}

\def\ket#1{{|#1\rangle}}
\def\bra#1{{\langle#1|}}
\def\4#1{{\boldsymbol{#1}}}
\def\8#1{{\widetilde{#1}}}

\def\mean#1{{\langle#1\rangle}}

\begin{document}

\title{Generalized Quantum Telecloning}

\author{Goren Gordon}
\affiliation{Department of Chemical Physics, Weizmann Institute of
Science, Rehovot 76100, Israel}
\author{Gustavo Rigolin}
\email{rigolin@ifi.unicamp.br} \affiliation{Departamento de
F\'{\i}sica da Mat\'eria Condensada, Instituto de F\'{\i}sica Gleb
Wataghin, Universidade Estadual de Campinas, C.P.6165, cep
13084-971, Campinas, S\~ao Paulo, Brazil}

\begin{abstract}
We present a generalized telecloning (GTC) protocol where the
quantum channel is non-optimally entangled and we study how the
fidelity of the telecloned states depends on the entanglement of
the channel. We show that one can increase the fidelity of the
telecloned states, achieving the optimal value in some situations,
by properly choosing the measurement basis at Alice's, albeit
turning the protocol to a probabilistic one. We also show how one
can convert the GTC protocol to the teleportation protocol via
proper unitary operations.
\end{abstract}

\maketitle

\section{Introduction}

Since the appearance of the quantum teleportation protocol
\cite{ben93} and its experimental demonstration
\cite{bou97,bos98}, whereby an arbitrary state describing a
quantum system can be transferred from one recipient (Alice) to
another (Bob), several new quantum communication protocols have
appeared. They allow the sharing of quantum states among several
recipients \cite{Lan04}, the sharing of quantum secrets
\cite{Hil99,Kar99,Gis01}, or the teleportation of an arbitrary
quantum state to many recipients, i.e. quantum telecloning
\cite{mur99}. The latter protocol does not violate the no-cloning
theorem \cite{woo82} since the fidelity of the telecloned states
with respect to the original one are not perfect, and decreases
with the number of copies. An optimal quantum telecloning protocol
has been presented in Refs. \cite{mur99,gis97} for two-level
systems (qubits) and later on quantum telecloning has been
demonstrated experimentally for continuous variables systems
\cite{cer85,koi06}.

These protocols are essential to many quantum information tasks
which require a secure transmission of quantum states. One example
is quantum information networks \cite{Cir97,Lan04}, which are
built of nodes in which quantum states are created, manipulated,
and stored. These nodes are connected by multipartite entangled
quantum channels and by properly using one or several of the
aforementioned protocols one could avoid errors and eavesdropping
during the transmission of a state between nodes
\cite{Lan04,Lo99}.

However, most treatments of these protocols assume bipartite or
multipartite maximally entangled channels, whereas in realistic
scenarios decoherence and noise ensure that that is not the case.
One suggested solution is quantum distillation protocols
\cite{Ben96}, which allow us to obtain a maximally entangled state
from a large ensemble of partially entangled states, although only
asymptotically. Another one is to dynamically control the
decoherence of the channel qubits \cite{gor06a,gor06b}.

In Ref. \cite{gor06c}, inspired by Ref. \cite{aga02}, and in Ref.
\cite{gor06d}, we have generalized the teleportation \cite{gor06c}
and quantum state sharing \cite{gor06d} protocols to an arbitrary
number of input qubits and shown that one can overcome the
fidelity decrease due to non-maximally entangled channels on
expense of transforming the protocols to probabilistic ones. These
generalized protocols give the parties freedom to allocate the
channel's resources to a continuous distribution between the
fidelity of the protocol and its probability of success to achieve
a given fidelity. Other interesting approaches using pure non-maximally
entangled resources were presented in Refs.
\cite{Guo00,Buz05,Kur06}. In Ref. \cite{Guo00} it was shown how to directly
teleport a qubit using non-maximally entangled pure channels. Contrary to
Ref. \cite{aga02}, in Ref. \cite{Guo00} Bob needs to implement a unitary
operation on his qubit and an ancillary plus a measurement on the ancillary
in order to finish the protocol. In Ref. \cite{Buz05} it was
discussed how to implement entanglement swapping using non-maximally pure
entangled states and in Ref. \cite{Kur06} how to construct an oblivious remote
state preparation procedure using non-maximally entangled resources.

In this contribution we present the generalized telecloning
protocol (GTC), where we generalize the standard quantum
telecloning protocol to non-optimally entangled multipartite
channels (see Fig.~\ref{Fig-schematic}). For a comprehensive
review of other interesting extensions of the telecloning protocol
see Ref. \cite{Ghi03}. By treating each qubit's degraded
contribution to the entanglement of the channel separately, we
show that one can overcome the resulting fidelity decrease by
applying appropriate modifications to the protocol. Our main
results show that: (a) the port's qubit influence on the
entanglement of the channel can be overcome by changing the
measurement basis; (b) the ancillary qubit's behavior has no
effect on the telecloned fidelity; (c) the copy qubits' behavior
has a non trivial influence on the fidelity of the telecloned
states and we show the optimal strategy to maximize the efficiency
of the protocol; and (d) it is possible to convert the GTC to the
generalized teleportation protocol (GTP) if one allows Alice to
implement certain types of unitary operations on the channel's
qubits.

\section{General formalism}

\begin{figure}[!ht]
\centering
\includegraphics[width=2.5in]{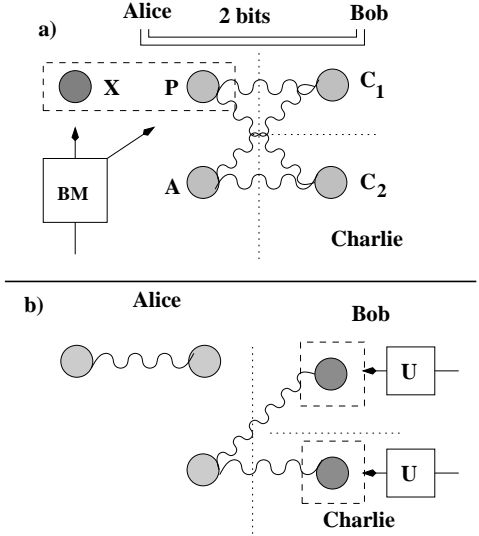}
\caption{\label{Fig-schematic} Alice performs a Bell measurement
(BM) on the qubit to be telecloned (X) and on the port qubit (P).
She then tells Bob and Charlie her measurement result (2 bits).
The copies $C_1$ and $C_2$ are then subjected to a proper unitary
operation (U). Note that waves represent the existence of pairwise
entanglement among the qubits and that the ancillary qubit (A) is
entangled with the copies at the end of the protocol.}
\end{figure}

We focus our attention on the ``$1\rightarrow 2$ quantum
telecloning", i.e. one original qubit and two copies. Let us
assume that Alice wishes to teleclone her state to Bob and
Charlie. The quantum channel used for the optimal telecloning
protocol \cite{mur99,gis97} is composed of four qubits, namely
port qubit, ancillary qubit and two copy qubits. The port and
ancillary qubits are assumed to be with Alice, although the
ancillary is not required to be there \cite{mur99}. One copy qubit
is with Bob while the other one is with Charlie
(Fig.~\ref{Fig-schematic}).

The channel state is given by:
\be \label{channel-def} \ket{\psi}_{PAC}=\frac{1}{\sqrt{2}}
\left(\ket{0}_P\otimes\ket{\phi_0}_{AC}+\ket{1}_P\otimes\ket{\phi_1}_{AC}\right),
\ee
where
\bea \label{phi0-def}
\ket{\phi_0}_{AC}&=&\sum_{j=0}^1\alpha_j\ket{\{0,1-j\},\{1,j\}}_A\nonumber
\\ &\otimes& \ket{\{0,2-j\},\{1,j\}}_{C},\\
\label{phi1-def}
\ket{\phi_1}_{AC}&=&\sum_{j=0}^1\alpha_j\ket{\{0,j\},\{1,1-j\}}_A \nonumber \\
&\otimes&\ket{\{0,j\},\{1,2-j\}}_{C},\\
\alpha_j&=&\sqrt{(2-j)/3}.
 \eea
Here the subscripts denote the port ($P$), ancillary ($A$) and
copies ($C$: $C_1$ with Bob and $C_2$ with Charlie). The state
$\ket{\{0,M-j\},\{1,j\}}$ represents the symmetric and normalized
state of $M$ qubits in which $M-j$ of them are in the state
$\ket{0}$ and $j$ are in the orthogonal state $\ket{1}$ (See ref.
\cite{mur99}). For $M=2$ we have explicitly,
\begin{eqnarray*}
&&\ket{\phi_0}_{AC} = \sqrt{\frac{2}{3}}|000\rangle_{AC} +
\sqrt{\frac{1}{6}}|101\rangle_{AC} +
\sqrt{\frac{1}{6}}|110\rangle_{AC}, \\
&&\ket{\phi_1}_{AC} = \sqrt{\frac{2}{3}}|111\rangle_{AC} +
\sqrt{\frac{1}{6}}|001\rangle_{AC} +
\sqrt{\frac{1}{6}}|010\rangle_{AC}.
\end{eqnarray*}

We analyze the influence of each qubit on the entanglement of the
channel by applying a qubit-specific `disentanglement' operator:
\be \label{disentanglement-def}
\hat{D}_i\left(\alpha\ket{0}_i\ket{\psi_0}+\beta\ket{1}_i\ket{\psi_1}\right)=
\frac{\alpha\ket{0}_i\ket{\psi_0}+n_i\beta\ket{1}_i\ket{\psi_1}}{\sqrt{|\alpha|^2+
|n_i\beta|^2}}, \ee
where $n_i$ can be complex and $|\alpha|^2+|\beta|^2=1$. For
example, when this operator is applied on a Bell state, e.g.
$\ket{\Phi^+}=1/\sqrt{2}\left(\ket{00}+\ket{11}\right)$, it
produces a non-maximally entangled state,
$\hat{D}_1(\ket{\Phi^+})=1/\sqrt{1+|n_1^2|}\left(\ket{00}+n_1\ket{11}\right)$.
When it is applied to the second qubit of the $W$ state,
$|W\rangle = (1/\sqrt{3})(|001\rangle + |010\rangle +
|100\rangle)$, we get
$\hat{D}_2(\ket{W})=1/\sqrt{2+|n_2^2|}\left(\ket{001}+n_2\ket{010}
+ \ket{100}\right)$. In other words, the application of
$\hat{D}_i$ on a state changes the $i-th$ qubit according to the
following map: $|0\rangle_i \rightarrow |0\rangle_i$ and
$|1\rangle_i \rightarrow n_i|1\rangle_i$. Note that the final
state is obtained normalizing the state obtained after we apply
the map.

It is worth mentioning that we called the map described in the
previous paragraph a `disentanglement' operator because the state
obtained after its application on a given maximally entangled
state does not have the same amount of entanglement as before. We
have a decrease on the entanglement content of the original state.
It is in this sense that one should understand this terminology.

Applying this operator on each qubit in the telecloning channel
results in:
%
\bea \ket{\psi;\{n\}}_{PAC}&=&A
\Big(\ket{0000}+\frac{n_Pn_{C_1}}{2}\ket{1010}+\frac{n_An_{C_1}}{2}\ket{0110} \nonumber \\
&& +\frac{n_Pn_{C_2}}{2}\ket{1001} + \frac{n_An_{C_2}}{2}\ket{0101}\nonumber \\
&&+n_Pn_An_{C_1}n_{C_2}\ket{1111}\Big)_{PAC},
\label{psi-channel-n} \eea
where
\bea A&=&\Big(1+\frac{|n_Pn_{C_1}|^2}{4}+\frac{|n_An_{C_1}|^2}{4}
+\frac{|n_Pn_{C_2}|^2}{4} +\frac{|n_An_{C_2}|^2}{4}\nonumber
\\
&&+|n_Pn_An_{C_1}n_{C_2}|^2\Big)^{-1/2}. \label{psi-norm}\eea
Note that $\{n\}=\{n_P,n_A,n_{C_1},n_{C_2}\}$ represents all the
`disentanglement' parameters.

Before we proceed we want to show how the entanglement of the
state (\ref{psi-norm}) depends on the values of $n_j$, where
$j=P,A,C_1$, and $C_2$. This analysis is important since it allows
one to connect the efficiency of the protocol to the entanglement
of the channel. Furthermore, it also justifies why we have called
$\hat{D}_i$ a `disentanglement' operator.

In order to quantify the entanglement of the channel we employ the
global entanglement $E_G^{(1)}$ proposed by Meyer and Wallach
\cite{mey02} and fully discussed and generalized in \cite{rig06},
\begin{equation}
E_G^{(1)}(\ket{\psi;\{n\}}_{PAC}) = 2 \left( 1 -
\frac{1}{4}\sum_{j=1}^{4}\mbox{Tr}(\rho_j^2)\right),
\end{equation}
where $\rho_j$ is the reduced density matrix describing qubit $j$,
obtained tracing out all qubits of the channel but $j$. One can
show \cite{rig06} that $E_G^{(1)}$ is the mean linear entropy of
the qubits belonging to the state (\ref{psi-norm}) and that it is
related to the purity of the qubits. For our purposes, $E_G^{(1)}$
is a fairly good multipartite entanglement quantifier
\cite{mey02,rig06}.

\begin{figure}[!ht]
\centering
\includegraphics[width=2.75in]{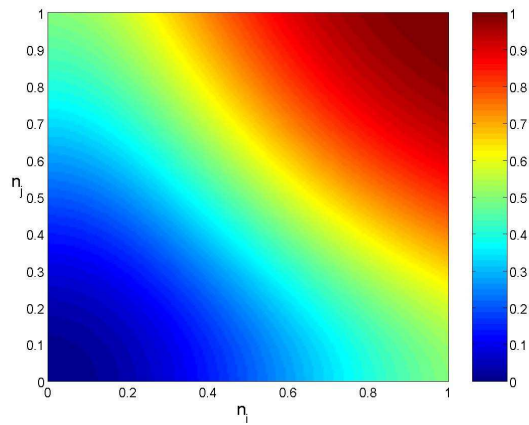}
\caption{\label{Fig_EG1_2D} Global entanglement, as given by Eq.
(\ref{EG1_2D}), as a function of $(n_i,n_j)=(n_1,n_2)=(n_A,n_P)$.}
\end{figure}

The general expression for $E_G^{(1)}$ is too cumbersome and not
insightful. Therefore, we show here the most representative cases
for $n_j$ real. Whenever all but one of the `disentangling'
parameters are equal to one, or, in other words, whenever we apply
$\hat{D}_i$ to only one of the channel's qubits we have,
\begin{equation}
E_G^{(1)}(\ket{\psi;n_j}_{PAC}) = \frac{1+6 n_j^2 +
n_j^4}{2(1+n_j^2) ^2}, \label{EG1}
\end{equation}
where $j=P, A, C_1$, or $C_2$. As depicted in Fig. \ref{CproEG1}
we see that the global entanglement is an increasing function of
$n_j$. When we deal with two free parameters, i.e. $n_i$ and $n_j$
different from one, we have two possibilities. For $(n_i,n_j)$ $=$
$(n_A,n_P)$ $=$ $(n_1,n_2)$ we get,
\begin{equation}
E_G^{(1)} = \frac{8 n_j^2 + n_j^4 + n_i^4(1 + 8n_j^4) + n_i^2 (8 +
38 n_j^2 + 8 n_j^4)}{2(2+n_j^2+n_i^2+2n_i^2n_j^2)^2}.
\label{EG1_2D}
\end{equation}
On the other hand, for $(n_i,n_j)$ $=$ $(n_A,n_1)$ $=$ $(n_A,n_2)$
$=$ $(n_P,n_1)$ $=$ $(n_P,n_2)$ we get
\begin{equation}
E_G^{(1)} = \frac{2(4 + 5 n_i^2 + n_j^2( 5 + (44 + 5 n_j^2) n_i^2
+ (5 + 4n_j^2)n_i^4 ))}{(5+n_j^2+n_i^2+5n_i^2n_j^2)^2}.
\end{equation}
Both expressions, however, have a similar behavior. Therefore, in
Fig. \ref{Fig_EG1_2D} we only plot Eq. (10). Note that, again, the
global entanglement is an increasing function of $n_i$ and $n_j$.

Let us now return to the telecloning protocol.  Using the channel
given in Eq.~(\ref{psi-norm}) Alice wants to teleclone an
arbitrary state, $\ket{\phi}_X=\alpha\ket{0}_X+\beta\ket{1}_X$, to
Bob and Charlie. The full initial state, with the qubit to
teleclone, is simply given by \be \label{initial state}
\ket{\phi}_{XPAC}=\ket{\phi}_X\otimes\ket{\psi;\{n\}}_{PAC}, \ee
and the protocol works as follows.

Alice performs a modified Bell measurement \cite{gor06c,gor06d},
i.e. she projects her original (X) and port (P) qubits onto the
following modified Bell basis: \bea
|\Phi^{+}_{m}\rangle &=& M (|00\rangle + m |11\rangle),\label{Bell1}\\
|\Phi^{-}_{m}\rangle &=& M (m^*|00\rangle - |11\rangle),\\
|\Psi^{+}_{m}\rangle &=& M (|01\rangle + m |10\rangle),\\
|\Psi^{-}_{m}\rangle &=& M (m^*|01\rangle -
|10\rangle),\label{Bell2} \eea where $M=1/\sqrt{1 +|m|^2}$. We
introduce, as will become clear soon, a free parameter ($m$) in
the protocol. It is a proper manipulation of this parameter that
allows Alice to overcome the fidelity decrease due to her port
qubit disentanglement ($|n_P|<1$). Each projective measurement
implemented by Alice on qubits $X$ and $P$ projects the ancillary
and copy qubits to the state $\ket{R_j}_{AC_1C_2}$, with
probability $P_j$. Here
$j=\{\Phi^+_m,\Phi^-_m,\Psi^+_m,\Psi^-_m\}$ stands for any
possible measurement result obtained by Alice.
Alice then sends Bob and Charlie her measurement result (two
bits). Then, both parties apply the appropriate unitary
transformation on their qubits,
$\{\Phi^+_m,\Phi^-_m,\Psi^+_m,\Psi^-_m\}\rightarrow
\{I,\sigma_z,\sigma_x,\sigma_z\sigma_x\}$. At the end of the
protocol Bob (Charlie) ends up with the state $\rho_{1(2),j}={\rm
Tr}_{A,C_{2(1)}}\left(\ket{R_j}_{AC_1C_2}\bra{R_j}\right)$, which
is obtained tracing out all but qubit $C_1(2)$. Therefore, Bob's
(Charlie's) fidelity for this run of the protocol is $F_{1(2),j} =
{}_X\bra{\phi}\rho_{1(2),j}\ket{\phi}_X$.

\section{Channel efficiency}

We now turn to estimate the efficiency of the protocol employing
the techniques developed in Ref. \cite{gor06c}. From now on
$\{n\}$ and $m$ are all real numbers since it can be shown that we
do not lose in generality by such assumptions \cite{gor06c}. In
general the probabilities $P_j$ and the fidelities $F_{1(2),j}$
depend on $\alpha$ and $\beta$. Moreover, Alice can change the
values of $\alpha$ and $\beta$ of the transferred state at will
for each run of the protocol. Therefore, in order to get $\alpha$-
and $\beta$-independent results we average over many
implementations of the protocol, i.e. over all possible pure state
inputs, obtaining the \textit{protocol efficiency} \cite{gor06c}
$$C^{pro}_{1(2)} = \sum_j \langle P_j F_{1(2),j}\rangle.$$

In the averaging process we will need the quantities $\langle
|\alpha|^2 \rangle$, $\langle |\alpha|^4 \rangle$, $\langle
|\beta|^2 \rangle$, $\langle |\beta|^4 \rangle$ and $\langle
|\alpha\beta|^2 \rangle$. In Ref. \cite{gor06c} they were shown to
be $\langle |\alpha|^2 \rangle = \langle |\beta|^2 \rangle = 1/2$,
$\langle |\alpha|^4 \rangle = \langle |\beta|^4 \rangle = 1/3$,
and $\langle |\alpha\beta|^2 \rangle = 1/6$. We can interpret
$C^{pro}$ as the average qubit transmission rate for a given
protocol choice \cite{gor06c}.

The averaged probabilities, Bob's average fidelities, and his
channel efficiency are: \bea
\mean{P_{\Phi^+_m}}&=&\mean{P_{\Psi^-_m}}=\frac{A^2M^2}{2}
\Big(1+\frac{n_P^2n_{C_1}^2m^2}{4}+\frac{n_A^2n_{C_1}^2}{4}\nonumber \\
&+&\frac{n_P^2n_{C_2}^2m^2}{4}+\frac{n_A^2n_{C_2}^2}{4}
+n_P^2n_A^2n_{C_1}^2n_{C_2}^2m^2\Big),\label{Prob1}\\
\mean{P_{\Phi^-_m}}&=&\mean{P_{\Psi^+_m}}=\frac{A^2M^2}{2}
\Big(m^2+\frac{n_P^2n_{C_1}^2}{4}+\frac{n_A^2n_{C_1}^2m^2}{4} \nonumber \\
&+&\frac{n_P^2n_{C_2}^2}{4}+\frac{n_A^2n_{C_2}^2m^2}{4}
+n_P^2n_A^2n_{C_1}^2n_{C_2}^2\Big),\eea
\bea \mean{F_{1,\Phi^+_m,\Psi^-_m}P_{\Phi^+_m,\Psi^-_m}}&=&
\frac{A^2M^2}{3}\Big(1+ \frac{n_A^2 n_{C_1}^2}{8}+ \frac{n_A^2
n_{C_2}^2}{4} \nonumber \\
&&+\frac{n_P n_{C_1}m}{2}+ \frac{n_P n_A^2 n_{C_1}n_{C_2}^2 m}{2}
\nonumber\\
&&+\frac{n_P^2 n_{C_1}^2 m^2}{4}+ \frac{n_P^2 n_{C_2}^2 m^2}{8}
\nonumber \\&&+ n_P^2 n_A^2 n_{C_1}^2 n_{C_2}^2 m^2
\Big),\\
\mean{F_{1,\Phi^-_m,\Psi^+_m}P_{\Phi^-_m,\Psi^+_m}}&=&
\frac{A^2M^2}{3}\Big(m^2+ \frac{n_A^2 n_{C_1}^2 m^2}{8} \nonumber \\
&& + \frac{n_A^2 n_{C_2}^2 m^2}{4} + \frac{n_P n_{C_1}m}{2}
\nonumber \\
&& + \frac{n_P n_A^2 n_{C_1}n_{C_2}^2 m}{2} + \frac{n_P^2
n_{C_1}^2}{4}  \nonumber
\\
&&+ \frac{n_P^2 n_{C_2}^2}{8}+ n_P^2 n_A^2 n_{C_1}^2
n_{C_2}^2\Big), \eea
\be C^{pro}_1=\frac{2}{3}\left(1+\frac{1}{2}
\frac{\mathcal{F}(\{n\})}{\mathcal{G}(\{n\})}\right),
\label{ChanEff1} \ee
with
\begin{eqnarray}
\mathcal{F}(\{n\}) &=&
(1+n_P^2)(1+n_{C_1}^2)(1+n_A^2n_{C_2}^2)c(n_P)c(n_{C_1})c(m)
\nonumber \\
&&-(n_A^2n_{C_1}^2+n_P^2n_{C_2}^2), \\
\mathcal{G}(\{n\}) &=& (n_P^2+n_A^2)(n_{C_1}^2+n_{C_2}^2)
+4(1+n_P^2n_A^2n_{C_1}^2n_{C_2}^2).\nonumber \\
\end{eqnarray}
Here $c(n)=2n/(1+n^2)$ is the concurrence \cite{Woo98} of the
state $1/\sqrt{1+n^2}\left(\ket{00}+n\ket{11}\right)$. On the
other hand, Charlie's fidelities and his channel efficiency are
simply obtained by changing $n_{C_1}\leftrightarrow n_{C_2}$. For
the standard telecloning protocol $\{n\}=m=1$ and one obtains the
well-known result of $\mean{P_j}=1/4, \mean{F_{1(2),j}P_j}=5/24$,
and $C^{pro}_{1(2)}=5/6$, which is the optimal average fidelity
\cite{mur99}.

We now begin to study each qubit's disentanglement effect on the
channel efficiency $C^{pro}$. We investigate how the port,
ancillary and copies' disentanglement influence the overall channel
efficiency and how we can remedy the disentanglement effect as
modelled by Eq.~(\ref{disentanglement-def}).

\subsection{Port qubit treatment}

The first qubit we treat is the port. Applying the map giving in
Eq. (\ref{disentanglement-def}) only to the port qubit (i.e.
$n_A=n_{C_{1,2}}=1.0$) we get:
\be \label{c-pro-port}
C^{pro}_{1(2)}=\sum_j\mean{F_{1(2),j}P_j}=\frac{11}{18}
\left(1+\frac{4c(m)c(n_P)}{11}\right). \ee
Note that for $n_P=m=1$ we obtain $C^{pro}_{1(2)}=5/6$, the
original telecloning efficiency \cite{mur99}. Moreover, noting
that for this case the channel can be written as
\begin{equation}
\ket{\psi;\{n\}}_{PAC}=\frac{1}{\sqrt{1+n_P^2}}\left( \ket{0}_P
\ket{\phi_0}_{AC} + n_P \ket{1}_P \ket{\phi_1}_{AC}\right),
\end{equation}
it is evident to see that the same treatment as in the Generalized
Teleportation Protocol (GTP) \cite{gor06c} and the Generalized
Quantum State Sharing (GQSTS) \cite{gor06d} applies here. By
simply changing the measurement basis (adjusting a proper $m$) and
choosing the proper acceptable measurements one can either retain
unit probability of success with low fidelity ($m=1$, accepting
all results), or transform the protocol to a probabilistic one
with optimal fidelity ($5/6$). For example, by choosing $m=n_P$ we
recover probabilistically \cite{gor06c,gor06d} the noiseless
telecloning protocol \cite{mur99}. For this choice of $m$, only
$\ket{\Phi^-_m}$ and $\ket{\Psi^+_m}$ are acceptable results both
of which furnishing the optimal fidelity for a given run of the
protocol (no need for averaging) \cite{gor06c,gor06d}.

\begin{figure}[!ht]
\centering
\includegraphics[width=2.75in]{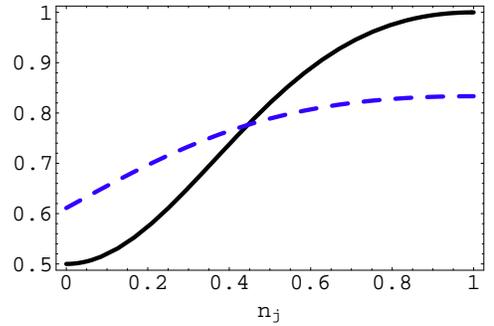}
\caption{\label{CproEG1} Global entanglement (black-solid), as
given by Eq.~(\ref{EG1}), and the channel efficiency
(blue-dashed), Eq.~(\ref{c-pro-port}), as a function of $n_j=n_P$
for $c(m)=1$.}
\end{figure}

Finally, we can see that the greater the channel efficiency (Eq.
(\ref{c-pro-port})) the greater the channel global entanglement.
See Fig.~\ref{CproEG1}.

\subsection{Ancillary qubit treatment}

Applying the map given in Eq. (\ref{disentanglement-def}) only to
the ancillary qubit (i.e. $n_P=n_{C_{1,2}}=1.0$) we get
\be \label{c-pro-ancilla}
C^{pro}_{1(2)}=\sum_j\mean{F_{1(2),j}P_j}=\frac{11}{18}
\left(1+\frac{4c(m)}{11}\right).
 \ee
It is interesting to note that the ancillary disentanglement ($n_A
< 1$) has no effect on the overall channel efficiency. In other
words, Eq.~(\ref{c-pro-ancilla}) does not depend on $n_A$. It is
worth noting that Eq.~(\ref{c-pro-ancilla}) is equal to
Eq.~(\ref{c-pro-port}) when $c(n_P)=1$, i.e., when one still has a
maximally entangled channel ($E_G^{(1)}=1$). Again we find that
the overall channel efficiency is optimal for $m=1$, namely
$C^{pro}_{1(2)}=5/6$.

\subsection{Copy qubit treatment}
The last case to consider is the one in which we apply the map to
the copies. In this case, we assume that the port and the
ancillary qubits are not affected (i.e. $n_P=n_A=1.0$). The
channel efficiency can be rewritten as
\bea \label{c-pro-copy} &C^{pro}_{1}=\frac{1}{2}\left( 1 +
\frac{2}{3}\left(\kappa^{(1)}+\kappa^{(2)}c(n_{C_1})c(m)\right)\right),\\
\label{symmetric-mutual-concurrence}
&\kappa^{(1)}=\frac{1}{1+\lambda}, \quad\kappa^{(2)}=\frac{1}{1+1/\lambda},\\
&\lambda=\frac{(1+n_{C_1}^2)(1+n_{C_2}^2)}{1+n_{C_1}^2n_{C_2}^2}.\label{lambda}
 \eea
For the second copy, $C^{pro}_{2}$ is given by changing
$n_{C_1}\leftrightarrow n_{C_2}$. As we discuss below,
Eq.~(\ref{c-pro-copy}) allows us to derive a couple of interesting
properties for this particular protocol. Firstly, let us analyze
some trivial limiting cases. For $m=1$, note that when
$n_{C_1}=n_{C_2}=1$ we obtain, as it should be,
$C^{pro}_{1(2)}=5/6$, the noiseless optimal limit. Moreover, when
$n_{C_1}=n_{C_2}=0$ we get $C^{pro}_{1(2)}=2/3$. This value can be
understood noting that for this case the channel is
$\ket{\psi;\{n\}}_{PAC}=\ket{0000}_{PAC}$, i.e. we have no
entanglement whatsoever. Thus the telecloning protocol can be seen
as the usual teleportation protocol whose efficiency is at most
$2/3$ when we have pure but not entangled channels \cite{gor06c}.
Furthermore, only for the case when $n_{C_1}=1$, we see that the
channel efficiency of the first copy does not depend on $n_{C_2}$,
as can be seen looking at Eq.~(\ref{lambda}). A similar argument
applies for the second copy channel efficiency. This is remarkable
and it means that the application of the map on the second (first)
copy changes the protocol efficiency of the first (second) copy in
a way that depends on the application of the map on the first
(second) copy. Finally, when $n_{C_2}=1$ one recovers
$C^{pro}_1=\frac{11}{18}(1+\frac{4}{11}c(m)c(n_{C_1}))$, similar
to Eq.~\eqref{c-pro-port}, with $n_P\leftrightarrow
n_{C_1}$. 
This shows that the action of the map on the port qubit ($n_P<1$)
changes $C^{pro}_1$ in exactly the same way as when the map acts
on just the first copy ($n_{C_1}<1$). However, in contrast to the
case where the map acts only on the port qubit, we were not able
to devise a procedure by which we can increase the fidelity of the
copies, even in a probabilistic protocol. In other words, equating
$m=n_{C_1}$ does not improve the fidelity of the copies, contrary
to a similar successful strategy ($m=n_P$) employed for the port
qubit case.

\begin{figure}[!ht]
\centering
\includegraphics[width=2.75in]{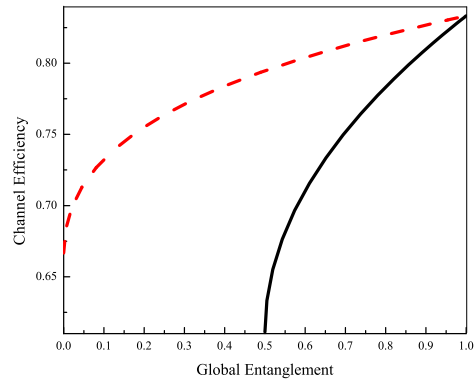}
\caption{\label{CproCopy1_EG1} Channel efficiencies as a function
of global entanglement when the `disentanglement' map is applied
to the port (black-solid) and to the copies (red-dashed), as given
by Eqs.~(\ref{c-pro-port}) and~(\ref{c-pro-copy})
($n_{C_1}=n_{C_2}$ or $n_A=n_P$), respectively. In all cases
$c(m)=1$.}
\end{figure}

We end this section showing that the channel efficiency is a
monotonic increasing function of the global entanglement, as
depicted in Fig.~(\ref{CproCopy1_EG1}).

\section{GTC to GTP conversion}

We end this article showing how one can convert the GTC to the GTP
protocol. In other words, we want to show how it is possible,
using first local and then global unitary operations, to convert
the GTC channel $\ket{\psi;\{n\}}_{PAC}$
(Eq.~(\ref{psi-channel-n})) to the GTP channel
$|\Psi^{GTP}_{n_{C_1}}\rangle$ $=$
$\left(1/\sqrt{1+n^2_{C_1}}\right)$ $(|00\rangle$ $+$ $n_{C_1}$
$|11\rangle)$. We want, therefore, to create a GTP channel between
Alice and copy 1 (Bob) in detriment of copy 2 (Charlie), who will
have a considerable decrease of his channel efficiency. This can
be achieved by `disentangling' copy 2 from Alice's qubit and copy
1. The final goal is to concentrate all the entanglement of the
channel between Alice and Bob.

\subsection{Local unitary operations}

Firstly, let us restrict ourselves to local unitary operations
(Alice's site). If we remember that the ancillary qubit (A) is
assumed to be with Alice, she can only operate on the port (P) and
ancillary qubits (See Fig. \ref{Fig-schematic}). An optimal
strategy for Alice, when we set $n_P=n_A=1$, $n_{C_2}=0$, and
$m=1$ for the measurement basis, is the application of the
following unitary operation on A and P:
\bea
\4R_{jk}(q)&=&\left(%
\begin{array}{cccc}
  1 & 0 & 0 & 0 \\
  0 & A_q & -qA_q & 0 \\
  0 & q^*A_q & A_q & 0 \\
  0 & 0 & 0 & 1 \\
\end{array}%
\right),\\
A_q&=&\frac{1}{\sqrt{1+|q|^2}}, \eea where $j,k=P,A$ are the two
qubits Alice acts upon. Here $\4R_{jk}$ is written in the basis
$\{\ket{00},\ket{01},\ket{10},\ket{11}\}$ and it is basically a
rotation in the $\ket{01},\ket{10}$ plane. The best result
(maximal channel efficiency) is achieved for the case $q=1$.
%
This choice for $q$ gives the channel (note the order in which the
qubits are written),
\bea |\Psi^{GTP}_{n_{C_1}/\sqrt{2}}\rangle_{PC_1AC_2}&=&
\frac{1}{\sqrt{1+n_{C_1}^2/2}}\nonumber \\
&& \times \left(\ket{00} +
\frac{n_{C_1}}{\sqrt{2}}\ket{11}\right)\!\!\otimes\! \ket{00}.
\eea
This is a GTP-like channel between $P$ and $C_1$ but with
$n_{C_1}/\sqrt{2}$ instead of $n_{C_1}$, which is the cost one
pays for the inaccessibility to the copy qubits. However, the
channel efficiency is still large since for $n_{C_1}=1$ we have
$C^{pro}_1=(6+2\sqrt{2})/9\approx 0.981$. Furthermore, we can also
implement with the above channel a probabilistic teleportation
protocol. This means we can have, sometimes, a unity fidelity
teleported state, i.e. a successful run of the protocol
\cite{gor06c,aga02}.

Borrowing from the case of $n_{C_2}=0$ and any $n_{C_1}$, to the
case of $n_{C_1}=1$ and $n_{C_2}<1$, we can make the same
transformations as before and arrive at the following channel
efficiencies: \bea
&&C^{pro}_1=\frac{6+2\sqrt{2}+5n_{C_2}^2}{9\,(1+n_{C_2}^2)}, \label{c-pro-bob}\\
&&C^{pro}_2=\frac{5+2\sqrt{2}n_{C_2}+6n_{C_2}^2}{9\,(1+n_{C_2}^2)}
\label{c-pro-charlie}. \eea
Looking at Eqs.~(\ref{c-pro-bob}) and (\ref{c-pro-charlie}) we can
draw several interesting conclusions: (i) $C^{pro}_1>C^{pro}_2$
for all $n_{C_2}$, which is a consequence of the fact that Alice's
qubit is more entangled with copy 1 qubit, located at Bob's, in
comparison with copy 2 at Charlie's; (ii) For $n_{C_2}=1$ we get
$C^{pro}_1<5/6$ and $C^{pro}_2<5/6$, showing that the unitary
transformation reduces the channel efficiency of the GTC protocol
when compared with the efficiency of a maximally entangled GTC
channel; (iii) Eqs.~(\ref{c-pro-bob}) and (\ref{c-pro-charlie}),
however, show that for $n_{C_2}\le\sqrt{\frac{4\sqrt{2}-3}{5}}$
one can achieve $C^{pro}_1\ge5/6$, thus highlighting the
transition point from the GTC to the GTP scenario.

\subsection{Global unitary operations}

If we now allow Alice to implement global unitary operations,
i.e., she has access, in addition to the port and ancillary
qubits, to at least one of the copies, she is able to recover the
GTP channel from the GTC channel via two transformations. We also
assume, from now on, that Alice has access only to copy 1, being,
thus, impossible for her to work with copy 2.

As we did before, we first assume that $n_P=n_A=1$, $n_{C_2}=0$,
and $m=1$ for the measurement basis. With this choice, the GTC
channel reads,
\bea \ket{\psi;\{n\}}_{PAC_1C_2}&=&\frac{1}{\sqrt{4+2n_{C_1}^2}}
\Big(2\ket{000} +n_{C_1}\ket{101}\nonumber \\
&& + n_{C_1}\ket{011}\Big)\otimes \ket{0}. \eea

First Alice implements the following unitary operation on the
ancillary and copy 1 qubits, setting $q=n_{C_1}/2$,
\bea
\4T_{jk}(q)&=&\left(%
\begin{array}{cccc}
  A_q & 0 & 0 & qA_q \\
  0 & 1 & 0 & 0 \\
  0 & 0 & 1 & 0 \\
  -q^*A_q & 0 & 0 & A_q \\
\end{array}%
\right),\\
A_q&=&\frac{1}{\sqrt{1+|q|^2}}, \eea where $j,k$ are the two
qubits Alice acts upon and now $\4T$ is basically a rotation in
the $\ket{00},\ket{11}$ plane. The resulting state,
$|\Psi^{(1)}\rangle=
\4T_{A,C_1}\left(\frac{n_{C_1}}{2}\right)\ket{\psi;\{n\}}_{PAC_1C_2}$,
is,
\bea
|\Psi^{(1)}\rangle_{PAC_1C_2}&=&\frac{1}{\sqrt{{4+2n_{C_1}^2}}}
 \left(\sqrt{4+n_{C_1}^2}\ket{0000} \right. \nonumber \\
&&\left. + n_{C_1}\ket{1010}\right). \eea
The second transformation Alice implements are on the port and
copy 1 qubits with
$q=n_{C_1}(1-\sqrt{4+n_{C_1}^2})/(n_{C_1}^2+\sqrt{4+n_{C_1}^2})$.
The final state,
$$
|\Psi^{GTP}\rangle_{PC_1AC_2}=
\4T_{P,C_1}\left(\frac{n_{C_1}(1-\sqrt{4+n_{C_1}^2})}{n_{C_1}^2+
\sqrt{4+n_{C_1}^2}}\right)|\Psi^{(1)}\rangle,
$$
is given as,
\be |\Psi^{GTP}\rangle_{PC_1AC_2}=\frac{1}{\sqrt{1+n_{C_1}^2}}
\left(\ket{00}+n_{C_1}\ket{11}\right)\otimes |00\rangle. \ee
This is exactly the GTP channel \cite{aga02,gor06c} we were
looking for. Therefore, if Alice has also access to copy 1, it is
possible to go from GTC to GTP.

Again, borrowing from the case in which $n_{C_2}=0$ and $n_{C_1}$
is the free parameter, to the case of $n_{C_1}=1$ and $n_{C_2}<1$,
we can make the same transformations as before and arrive at the
following channel efficiencies:
\bea
&&C^{pro}_1=\frac{135+77n_{C_2}^2}{135(1+n_{C_2}^2)},\\
&&C^{pro}_2=
\frac{135+(8\sqrt{5}+159)n_{C_2}^2+24\sqrt{15}n_{C_2}}{270(1+n_{C_2}^2)}.
\eea
Here, again, we have the following interesting results: (i)
$C^{pro}_1>C^{pro}_2$ for all $n_{C_2}$, reflecting the
concentration of entanglement between port and copy 1; (ii) For
$n_{C_2}=1$ we get $C^{pro}_1<5/6$ and $C^{pro}_2<5/6$, showing
that the transformations also reduce the channel efficiency of the
GTC protocol when compared with the efficiency for the maximally
entangled GTC channel; (iii) Finally, manipulating $C^{pro}_1$,
one sees that for $n_{C_2}\le\sqrt{45/71}$ one can achieve
$C^{pro}_1\ge5/6$, thus showing the transition point from the GTC
to the GTP scenario.

\section{Experimental proposal}

The main experimental challenge in order to implement the GTC is
the ability of Alice to apply on her qubits a generalized Bell
measurement. In other words, Alice must project the port qubit (P)
and the one to be telecloned (X) onto one of the four generalized
Bell states given in Eqs. (\ref{Bell1})-(\ref{Bell2}).
Fortunately, this can be achieved for the following qubit
encodings \cite{Kim04}: (i) single-photon state and the vacuum
state; (ii) a vertically and a horizontally polarized photon
state; and (iii) two coherent light states with opposite phases.
Using linear optical schemes Kim \textit{et al.} \cite{Kim04} have
shown how one is able to implement a generalized Bell measurement
for each one of the above three possible qubit encodings. For the
first two encodings, not all generalized Bell states can be
distinguished via linear optics, although the last one allows an
almost perfect discrimination among the four generalized Bell
states.

\section{Conclusion}

To conclude, we have shown that decreasing the entanglement of the
quantum channel needed for a perfect quantum telecloning protocol
results in non-trivial protocol efficiencies which depend on the
specific mechanism used to decrease its entanglement content
(`disentanglement' process). We have analyzed all the three
possible `disentanglement' scenarios. Firstly, acting locally on
the port qubit, the reduction of the channel's entanglement can be
dealt with in a probabilistic manner, similar to the approach
employed for the generalized teleportation and quantum state
sharing protocols. Here we can achieve the optimal fidelity for
the telecloned qubits by properly rotating Alice's measurement
basis. Secondly, the ancillary's disentanglement has no effect on
the overall average efficiency, as expected from an ancillary.
Thirdly, the copies' disentanglement cannot be counter attacked
using the port's disentanglement approach, i.e., there is no
rotation on Alice's measurement basis allowing, even
probabilistically,  the optimal fidelity for both telecloned
qubits. Finally, we have also shown how one can convert the
generalized telecloning channel, either using local or global
unitary operations, to the generalized teleportation channel. All
these results highlight that non-maximally pure entangled channels
can also be employed to the direct implementation of quantum
telecloning, although only probabilistically. And this suggests
that a promising route for further analysis is the study of what
can be done probabilistically using directly, i.e. without
distillation protocols, non-maximally mixed entangled channels.

\begin{acknowledgments}
GR thanks Funda\c{c}\~ao de Amparo \`a Pesquisa do Estado de
S\~ao Paulo (FAPESP) for funding this research.
\end{acknowledgments}

\end{document}